# Photo-Induced Ultrafast Symmetry Switch in SnSe

Yadong Han[1,2,#], Junhong Yu[1,2,#], Hang Zhang[1,2], Fang Xu[1], Kunlin Peng[3], Xiaoyuan Zhou[3], Liang Qiao[4], Oleg V. Misochko[1,5], Kazutaka G. Nakamura[6], Giovanni M. Vanacore,[7,*] Jianbo Hu[1,2,*]

[1] *State Key Laboratory for Environment-Friendly Energy Materials, Southwest University of Science and Technology, Mianyang 621010, China*

[2] *Laboratory for Shock Wave and Detonation Physics, Institute of Fluid Physics, China Academy of Engineering Physics, Mianyang 621900, China*

[3] *College of Physics and Institute of Advanced Interdisciplinary Studies, Chongqing University, Chongqing 401331, China*

[4] *School of Physics, University of Electronic Science and Technology of China, Chengdu 610054, China*

[5] *Institute of Solid State Physics, Russian Academy of Sciences, 142432 Chernogolovka, Moscow region, Russia*

[6] *Materials and Structures Laboratory, Tokyo Institute of Technology, R3-10, 4259 Nagatsuta, Yokohama 226-8503, Japan*

[7] *Department of Materials Science, University of Milano-Bicocca, Via Cozzi 55, 20121, Milano, Italy*

[#]*These authors contribute equally*

[*]*To whom correspondence should be addressed. Email: jianbo.hu@caep.cn (JH); giovanni.vanacore@unimib.it (GMV)*

**Abstract:** Layered tin selenide (SnSe) has recently emerged as a high-performance thermoelectric material with the current record for the figure of merit (*ZT*) observed in the high-temperature *Cmcm* phase. So far, access of the *Cmcm* phase has been mainly obtained via thermal equilibrium methods based on sample heating or application of external pressure, thus restricting the current understanding only to ground-state conditions. Here, we investigate the ultrafast carrier and phononic dynamics in SnSe. Our results demonstrate that optical excitations can transiently switch the point-group symmetry of the crystal from *Pnma* to *Cmcm* at room temperature in a few hundreds of femtoseconds with an ultralow threshold for the excitation carrier density. This non-equilibrium *Cmcm* phase is found to be driven by the displacive excitation of coherent

$A_g$ phonons and, given the absence of low-energy thermal phonons, exists in SnSe with the status of 'cold lattice with hot carriers'. Our findings provide important insight for understanding non-equilibrium thermoelectric properties of SnSe.

Understanding the physical properties of functional materials has been mostly achieved under thermal equilibrium conditions via conventional strategies, such as: tuning the temperature[1–3], the pressure[4–6], or chemical doping[7–10]. With the development of pump-probe techniques based on femtosecond lasers, an ultrashort optical pulse can temporarily destroy ground state ordering and bring the material into an out-of-equilibrium state[11–17]. The investigation of nonequilibrium dynamics is of great interest for both deciphering the fundamental details of the transient evolution [18,19], and accessing 'hidden states' of materials characterized by unusual properties inaccessible under equilibrium ergodic conditions[20–22]. Ultrafast non-equilibrium states have been frequently observed in crystals characterized by symmetry-lowering Peierls distortion. These observations have been able not only to unravel new fascinating physical phenomena (e.g., ultrafast switching between different topological states) [23–25], but also to achieve unprecedented properties for practical device applications (e.g., accelerating the operation speed of phase change materials up to the THz range) [14,26].

SnSe was recently reported to achieve a *ZT* value (i.e., the dimensionless figure of merit to characterize the heat-to-power conversion efficiency) of approximately 2.6 at 923 K along a particular crystallographic direction (the *b* axis) with the lattice transformed in the high-temperature *Cmcm* phase, setting the benchmark for high thermoelectric performance[1,27–29]. As shown in **Figure 1a**, SnSe can crystallize into two phases with different point-group symmetry: a layered orthorhombic *Pnma* phase (where $a \neq c \neq b$) at the ambient temperature with prototypical Peierls distortions, and a high-symmetry *Cmcm* phase (where $a \neq b = c$) characterized by a giant anharmonicity able to destroy the thermal transport[30–32]. Traditional steady-state methods[1,2,33,34] that are used to access the *Cmcm* phase are mainly relying on thermally heating the system to trigger the symmetry switch at a temperature above 800 K. However, such methods are not able to access the correlation between electronic and lattice degrees of freedom, which instead fundamentally determines the energy transport in SnSe.

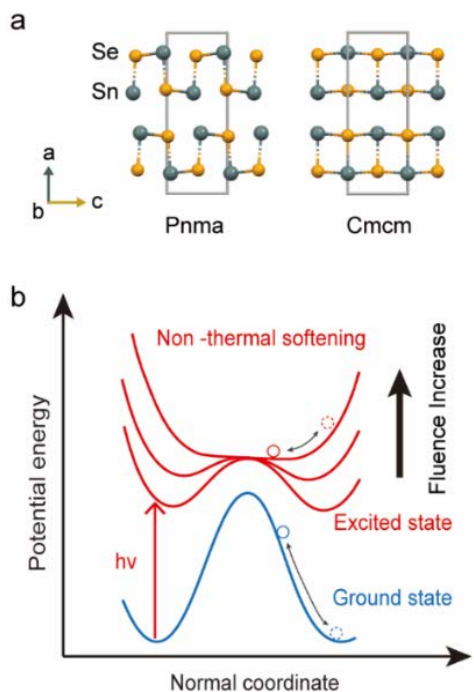

**Figure 1. Illustration of lattice structure and lattice softening in SnSe.** (a) The double bilayer structure of SnSe in the *Pnma* phase with Peierls distortions (left panel) and the high-symmetry *Cmcm* phase (right panel). The conventional unit cell is shown as a grey box. (b) Schematic of the potential energy surface of SnSe. During the thermal softening process (blue curve), the anharmonic vibrations in the ground state can only be achieved at high temperatures. Instead, under non-equilibrium conditions, an ultrafast photoexcitation promotes electrons from the ground state to the excited state (solid red curve). The energy exchanged with specific phonon modes can induce non-thermal softening and drive the lattice into the *Cmcm* phase, where atomic vibrations can reach the anharmonic region much easier than at the ambient condition.

To address this issue, we investigate ultrafast carrier and phononic dynamics in SnSe via femtosecond pump-probe spectroscopy. We observed an ultrafast photoinduced switch of point-group symmetry from *Pnma* to *Cmcm* in a few hundreds of femtoseconds with an ultralow carrier density threshold ($1.26 \times 10^{17}$ cm$^{-3}$). Based on the redshifted frequency and the increased anharmonic damping rate of coherent phonon modes, we determine that the displacive generation of coherent $A_g$ phonons is playing the leading role in the lattice softening. The transition to a nonequilibrium *Cmcm* phase is thus driven by an anti-Peierls distortion where the modification of the potential energy surface (from double-well to a single-well) is induced by the electronic excitation coupled to such coherent $A_g$ modes[35] (see **Figure 1b**). Given the absence of low-energy thermal phonons on the timescale of the transition, our experimental observation of a photoinduced point-group symmetry switch in SnSe provides an ideal platform to study thermoelectric properties at the status of 'cold lattice with hot carriers'.

We investigate excitation of coherent phonons[36–38] following an above bandgap excitation at ∼1.55 eV (the bandgap in SnSe is ∼1.30 eV[34,39,40]). The experiment adopts a standard optical pump-probe setup[37,38] in the reflection geometry to obtain the transient reflectivity induced by coherent phonon oscillations (more details are provided in **Figure S1**). **Figure 2a** presents the time-resolved reflectivity of freshly exfoliated SnSe samples (see synthesis details in **Methods**) with an excitation fluence of ~24 μJ/cm$^2$ at the ambient condition. There are two components in the photoinduced reflectivity transient trace: (1) at early times (< 100 fs), a non-oscillatory response due to excitation and relaxation of nonequilibrium carriers is observed, which will decay at about 100 fs; (2) at later times (> 100 fs), the transient reflectivity starts to oscillate in a sinusoidal manner caused by coherent excitation of the zone-center optical phonons.

To obtain transient information of coherent phonon modes under the non-equilibrium condition, we perform continuous wavelet transform (CWT)[41–43] to convert the transient reflectivity trace into the time-frequency chronogram. Note that the incoherent part (i.e., the non-oscillatory signal) is fitted by a sum of exponential functions and subtracted from the photoinduced reflectivity (see the description in **Figure S2**). As seen in **Figure 2b**, the time-frequency chronogram is dominated by a prominent peak with a frequency at ~67.5 cm$^{-1}$ and another weaker peak located at ∼29.7 cm$^{-1}$. Moreover, there are two additional feeble peaks located at ∼130.7 and ~147 cm$^{-1}$ (see the zoom-in spectra in **Figure S3**), whose integrated peak intensities are almost two orders of magnitude weaker than those of the dominant peaks. We found that the frequencies of these four coherent modes fairly agree as a whole – although slightly redshifted – with those obtained by spontaneous Raman scattering in a SnSe crystal[33,39,44]. Thus, these four coherent modes can be assigned to the four fully symmetric $A_g$ phonons (with the frequency from low to high): the $A_g^{(0)}$, $A_g^{(1)}$, $A_g^{(2)}$, and $A_g^{(3)}$ modes (normal vectors of these four modes are schematically shown in **Figure 2c**).

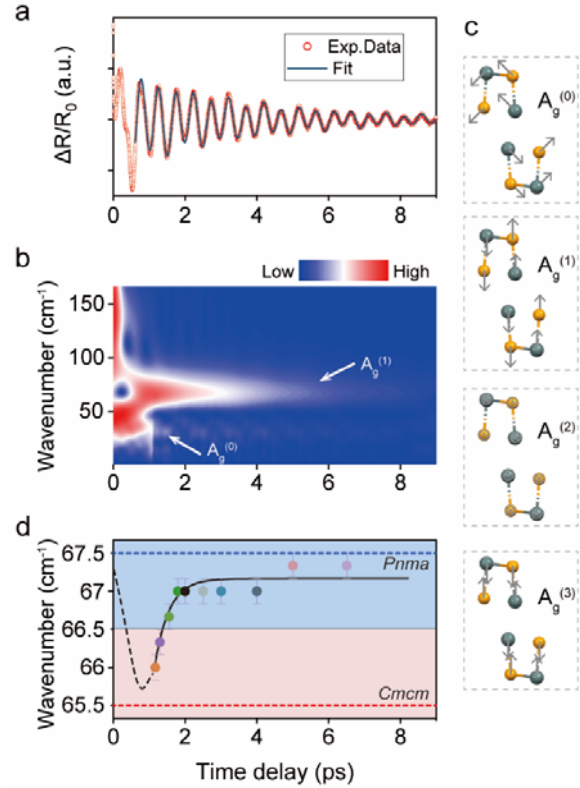

**Figure 2. Coherent phonon responses in photoexcited SnSe with an excitation fluence of 24 μJ/cm².** (a) The transient reflectivity changes ($\Delta R/R_0$) in SnSe at ambient conditions. The open circles are experimental data and the blue solid line is a least-squares fit. (b) The continuous wavelet transform (CWT) chronogram is obtained from the transient oscillation trace in Figure 2a. (c) Four fully symmetric oscillations can be resolved in the CWT chronogram. Their corresponding atomic displacements are presented. The traces are vertically offset for clarity. (d) The transient frequency of the coherent $A_g^{(1)}$ mode is extracted from the CWT chronogram, providing clear evidence of the point-group symmetry switch from *Pnma* to *Cmcm*. The boundary line between the red/blue shaded regions indicates the equilibrium *Pnma* to *Cmcm* transition frequency (~66.5 cm$^{-1}$) measured in temperature-dependent Raman spectroscopy[30]. The top and bottom dashed lines indicate the static frequency measured for *Pnma* at room temperature and *Cmcm* at 1000 K, respectively.

Owing to bond anharmonicity in *Cmcm* phase, peak frequencies of all $A_g$ phonons would redshift significantly when the anti-Peierls distortion occurs (i.e., soft phonons across a phase boundary between *Pnma* and *Cmcm*) [31]. Here we take the $A_g^{(1)}$ mode as an indicator of non-equilibrium dynamics, which assures a high signal-to-noise ratio ($A_g^{(2)}$ or $A_g^{(3)}$ mode are too weak to see a definite trend of the frequency shift) and allows to compare the transient phonon frequency with the results obtained in temperature-dependent Raman spectroscopy (the low-frequency $A_g^{(0)}$ mode is not resolvable in frequency-domain measurements)[33,45–47]. **Figure 2d** shows the transient peak-frequency shift of the $A_g^{(1)}$ mode extracted from **Figure 2b** (the separated phonon spectra at different delay times are also presented in **Figure S4**). In the early time window ($\Delta t < 1.5$ ps) the mode is located in the region (shaded in red) only observed in equilibrium Raman spectra when SnSe is heated into the *Cmcm* phase. Such an abrupt frequency blueshift that instantaneously crosses the frequency boundary at ~66.5 cm$^{-1}$ only at later times stabilizes back at the frequency of the initial *Pnma* phase. This is clear evidence of an ultrafast symmetry switch from *Pnma* to *Cmcm* under above band-gap optical excitation.

Further insight into the microscopic origin of the observed non-equilibrium dynamics comes from exploring the temporal evolution of the system at different excitation fluences. **Figure 3a** presents the photoexcitation-induced reflectivity change of SnSe for different selected pump fluences at the ambient condition. We have observed that at higher fluences the transient reflectivity undergoes a more noticeable frequency redshifting and a faster decay process. The redshifting trend is more clearly seen in the fast Fourier transform (FFT) spectra presented in **Figure 3b-c**. The $A_g^{(0)}$ phonon mode exhibits a larger deviation from the harmonic behavior (i.e., a more pronounced frequency softening, as large as 16%) with respect to the $A_g^{(1)}$ mode. This is a further evidence supporting our interpretation of a photoinduced switch of point-group symmetry in SnSe. In fact, the $A_g^{(0)}$ phonon is determined by in-plane atomic movements along the *c*-axis, whereas the $A_g^{(1)}$ phonon is generated by out-of-plane displacements (see **Figure 2c**). Only the coherently excited $A_g^{(0)}$ mode can significantly modulate the displacement of Sn atoms from off-centering sites and corrugate the bilayers along the *c*-direction. At certain excitation strength, the $A_g^{(0)}$ mode reduces the off-centering displacement to zero and changes the $C_{2v}$ symmetry of the layer to $D_{2h}$. When the Peierls distortion has been removed in SnSe, the high-symmetry *Cmcm* phase will thus display a more noticeable phonon softening in the low-frequency mode.

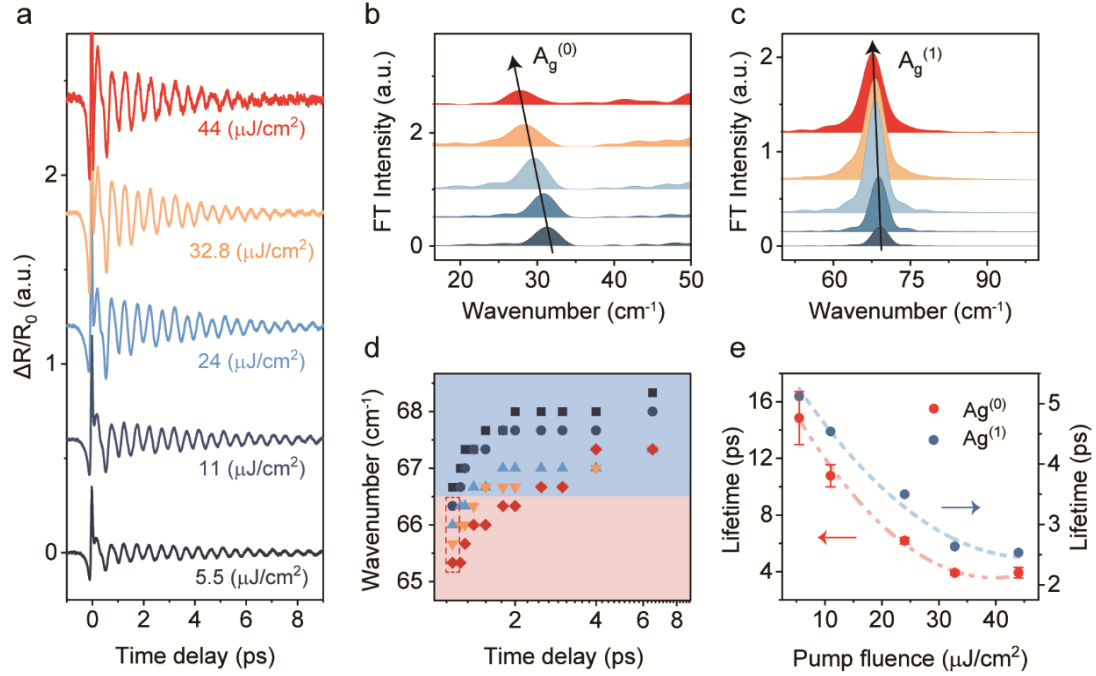

**Figure 3. Pump fluence dependence of coherent vibrational dynamics in SnSe.** (a) The transient-photoexcitation-induced reflectivity changes ($\Delta R/R_0$) in SnSe at the ambient condition are shown for different pump fluences. The traces are offset along the y-axis for clarity and corresponding fluences are indicated along each trace. Fast Fourier transform of the transient reflectivity after subtracting the non-oscillatory signal are shown to investigate the frequency shift of the $A_g^{(0)}$ mode (b) and the $A_g^{(1)}$ mode (c). The black arrow is a guide to the eyes. (d) The measured transient frequency change of the $A_g^{(1)}$ mode is shown for different pump fluences, highlighting a transition threshold of 11 μJ/cm². (e) The phonon lifetimes of the $A_g^{(0)}$ mode and the $A_g^{(1)}$ mode, as extracted by the combination of damped harmonic oscillations, are shown as a function of pump fluence.

By conducting a CWT analysis for all measured fluences, we identify a transition threshold around 11 μJ/cm² (see **Figure 3d** and the 2D chronogram for each fluence in **Figure S5**). To the best of our knowledge, this is the lowest electronic excitation required in Peierls-distorted crystals to reach a transient high-symmetry structure (other functional materials have been reported to require at least several mJ/cm² of excitation fluence[11,35,48,49]), confirming SnSe as a superior candidate for the design of thermoelectric devices.

Further insights into the softening of coherent phonon modes can be obtained by characterizing their lifetime, which has been extracted for the two modes of interest by modeling the transient reflectivity data considering the system as a combination of damped harmonic oscillators with the chirping effect[26,41,42]. As shown in **Figure 3e**, an increased fluence corresponds to a faster decay process of the $A_g^{(0)}$ and $A_g^{(1)}$ modes, implying an enhanced three-phonon scattering. Such increased phonon-phonon interaction thus leads to stronger anharmonic damping of the coherent lattice modes, and hence suppressed thermal transport. This observation represents an important signature of anharmonicity in the *Cmcm* phase under non-equilibrium conditions.

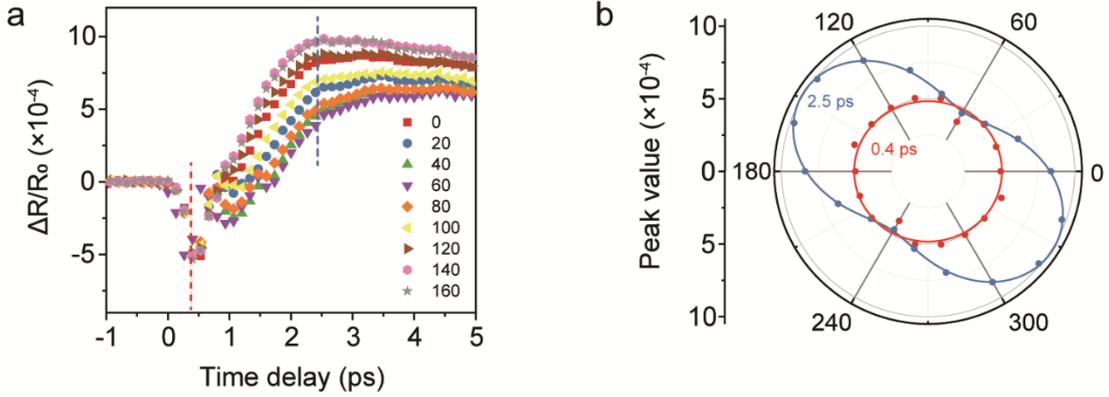

**Figure 4. Carrier dynamics in photoexcited SnSe with an excitation fluence of 24 μJ/cm².** (a) Polarization-dependent transient reflectivity profiles in SnSe. $\Delta R = R^{with\ pump} - R^{without\ pump}$ and $R^{without\ pump}$ always measures the reflection of probe pulse in *Pnma*-SnSe without pump pulse excitation. (b) Polar plots of the amplitude of reflectivity signal at two different delay times as marked by the dashed red and blue lines in panel a.

As additional evidence for the proposed microscopic transition mechanism, we have studied the carrier dynamics in SnSe by monitoring the ultrafast dielectric response of the sample measured via transient reflectivity experiments. First-principles calculations[31] showed that *Pnma*-SnSe exhibits an indirect bandgap while *Cmcm*-SnSe has a direct bandgap, suggesting that the latter has a stronger interband transition strength (i.e, a higher absorption coefficient) than the former. As a result, the transient reflectivity *ΔR/R* profiles should become negative when the SnSe sample is transiently driven into the *Cmcm* phase. A gradual change into the typical bleaching behavior would then occur in correspondence with the sample recovery toward the *Pnma* phase.

In **Figure 4a** we show the polarization-dependent transient reflectivity profiles measured at an excitation fluence of 24 μJ/cm², allowing us to directly access the dielectric response of the system. Here we can clearly distinguish at early times a negative band peaked at around 400 fs and lasting until about 1.2 ps, before turning into a positive trend at longer delay times. Such time scale agrees well with the frequency shifting observed via coherent phonon spectroscopy data shown in **Figure 2d**. We believe this is further evidence for the observed symmetry switch between *Pnma* and *Cmcm* phases.

Moreover, since these two phases have different structure symmetries, it is natural to expect for the dielectric response to show distinct polarization angle dependences at early times, when the SnSe is transiently driven into the *Cmcm* phase, with respect to later times, when the *Pnma* phase is recovered. To confirm this hypothesis, we have analyzed the polarization-dependent amplitude of the reflectivity signal at two different delay times (indicated by the two dashed lines in **Figure 4a**): the first one is selected at the peak of the negative band (0.4 ps) and the second one is located at ~ 2.5 ps. As shown in **Figure 4b**, the polar plot at ~ 2.5 ps shows a strong anisotropy, compatible with a low-symmetry *Pnma* phase. Instead, the polar plot at ~ 0.4 ps is almost isotropic, which is consistent with a higher-symmetry *Cmcm* phase characterized by a more isotropic band structure in different lattice axis.

The photoinduced generation of a *Cmcm* phase in SnSe with such a low excitation fluence (11 µJ/cm$^2$) opens a wide range of attractive possibilities for novel applications. For instance, one could use light, rather than external heat, to study energy transport in TE materials. Owing to the non-equilibrium conditions created by the femtosecond pulse, photo-induced hot carriers would live for few to tens of picoseconds before transferring their energy into incoherent low-energy thermal phonons. This would allow us to investigate the hot carriers' contribution to the energy transport within a relatively cold lattice, while simultaneously avoiding those degrading effects related to heating-based methods, such as material fatigue[50], irreversible material damage[51], and unwanted lattice distortion[52]. This could open new routes for the dynamic manipulation of electron-phonon coupling or even disentangling the electron/phonon contributions in energy transport. Finally, it is worth noticing that, although the time window for accessing the *Cmcm* phase is relatively short (about 1.2 ps in our experiments), one can extend it significantly by using multi-pulse excitation schemes[53,54].

In conclusion, we have investigated the ultrafast carrier and phononic dynamics of SnSe in this work. We have observed an ultrafast photoinduced transition from *Pnma* to *Cmcm* point-group symmetry evolving on a few hundreds of femtoseconds with an excitation fluence as low as 11 μJ/cm$^2$. The experimental results also demonstrate that the displacive excitation of coherent $A_g$ phonons is playing the leading role in the anti-Peierls distortions driving the symmetry switch. Such a mechanism is thus responsible for the strong anharmonicity of the lattice and reveals the non-thermal phonon softening nature of the symmetry switch. Our work, which demonstrates the ability to access the *Cmcm* phase in SnSe without thermally heating the lattice, would allow scientists in the field to go beyond the current thermal-equilibrium-based thinking and foster novel design ideas to improve the performance of thermoelectric materials.

# Methods

**Synthesis of SnSe crystal.** Sn granules (grain size: 1 - 6 mm, 99.999%), Se granules (grain size: 1 - 6 mm, 99.999%) were weighed according to the stoichiometry of SnSe, and loaded into cone-shaped silica tubes and sealed under 1 Pa. Then the cone-shaped silica tube is placed inside another large silica tube, again evacuated and flame-sealed, which aims at protecting the crystal from oxidation because the cone-shaped silica tubes are vulnerable as the crystal structure undergoes a phase transition from the Cmcm to Pnma phase. The quartz tubes were heated up to 1273 K over 12h, kept at that temperature for 20h, and then grown by a modified Bridgeman method. Excellent single-crystalline ingots of SnSe with a diameter of 13 mm and a height of 40 mm were readily obtained.

**Coherent phonon spectroscopy**. Ultrashort laser pulses with the duration of 45 fs and the photon energy of 1.55 eV (λ=800 nm) were delivered by a Kerr-lens mode-locked Ti: sapphire oscillator at the repetition rate of around 86 MHz. The laser acts both as a pump and a probe. Light polarization of pump and probe pulses is set to be orthogonal to each other to eliminate interference artifacts. The delay between the pump and probe is modulated by a fast scan at the frequency of 20 Hz to obtain the transient reflectivity change $\Delta R_{eo}(t)/R_0$, where $\Delta R_{eo}(t)$ is the difference between two orthogonal components of the reflected probe with pump, and $R_0$ is the reflectivity without optical excitations.

## Supporting Information

Experimental setup; data processing; oscillatory time-domain data and their corresponding Fourier transforms; the phonon spectra and corresponding FWHM at different time delays extracted from the CWT chronogram; the CWT chronogram for different excitation fluences.

## Acknowledgments

The authors thank Hiromu Matsumoto and Tsukasa Maruhashi for experimental assistance. This work was supported by Science Challenge Project (No. TZ2018001), National Natural Science Foundation of China (No. 11872058), Project of State Key Laboratory of Environment-friendly Energy Materials, Southwest University of Science and Technology (20fksy06), Russian Fund for Basic Research (No. 20-02-00231), Japan Society for the Promotion of Science KAKENHI (17H02797, 19K22141) and Collaborative Research Projects (CRP) -2021 of Laboratory for Materials and Structures (MSL), Tokyo Institute of Technology. G. M. V. acknowledges support from the SMART-electron project that has received funding from the European Union's Horizon 2020 Research and Innovation Programme under grant agreement No 964591.

## Author contributions

Y.H. and J.Y. contributed equally to this work. J.H. conceived the idea; X.Z. and K. P. prepared the samples; Y.H., H.Z., and K.N. performed the measurements, Y.H., O.M., J.Y., F.X., and J.H. analyzed the data; J.Y., Y.H., O.M., G.M.V., and J.H. wrote the manuscript with the input of all other authors.